\documentclass[prl,aps,twocolumn,preprintnumbers,superscriptaddress]{revtex4}
\usepackage{latexsym,epsfig,amssymb,amsfonts,amsmath,graphicx,color,bm,times,hyperref,amstext,epstopdf}

\usepackage{setspace}                

\usepackage{graphicx,color}          
\usepackage{epsfig}                  
\usepackage{wrapfig}
\usepackage{subfigure}
\usepackage{longtable}               
\usepackage{mathrsfs}                

\usepackage{amsthm}
\theoremstyle{plain}

\usepackage{dsfont}                  
\usepackage[a4paper]{anysize}        
\usepackage{braket}

\newtheorem{theorem}{Theorem}

\newtheorem{conj}{Conjecture}

\begin{document}
\title{A Shielding Property for Thermal Equilibrium States on the Quantum Ising Model}
\author{N. S. M\'oller}
\affiliation{Departamento de F\'isica, Universidade Federal de Minas Gerais,
Belo Horizonte, MG, Brazil}
\author{A. L. de Paula Jr}
\affiliation{Departamento de F\'isica, Universidade Federal de Minas Gerais,
Belo Horizonte, MG, Brazil}
\author{R. C. Drumond}
\affiliation{Departamento de Matem\'atica, Universidade Federal de Minas Gerais,
Belo Horizonte, MG, Brazil}

\date{\today}
\begin{abstract}
We show that Gibbs states of non-homogeneous transverse Ising chains satisfy a 
\emph{shielding} property. Namely, whatever the fields on each spin and exchange couplings between 
neighboring spins are, if the field in one particular site is null, the reduced states of the 
subchains to the right and to the left of this site are \emph{exactly} the Gibbs states of each 
subchain alone. Therefore, even if there is a strong exchange coupling between the extremal sites 
of each subchain, the Gibbs states of the each subchain behave as if there is no interaction between 
them. In general, if a lattice can be divided into two disconnected regions 
separated by an interface of sites with zero applied field, we can guarantee a similar 
result only if the surface contains a single site. Already for an interface with two sites we show 
an 
example where the property does not hold. When it holds, however, we show that if a perturbation 
of the Hamiltonian parameters is done in one side of the lattice, the other side is completely
unchanged, with regard to both its equilibrium state and dynamics.
\end{abstract}

\maketitle

\textit{Introduction.}---The transverse Ising chain is a paradigmatic model of quantum many body 
systems. It is exactly soluble via Jordan-Wigner transformation~\cite{LSM} and exhibits a quantum 
phase transition from a paramagnetic phase to a ferromagnetic one~\cite{IPT}. It is frequently used 
as a benchmark for analytical~\cite{ATec} or numerical techniques~\cite{Ntec}. It can be used to
illustrate or test new concepts, such as the whole of entanglement on phase 
transitions~\cite{EPT}, decoherence of open quantum systems~\cite{DOS} and quantum thermodynamics 
definitions of work~\cite{QTD}. It is also more than a toy model, being used to describe some 
trapped cold rubidium atoms~\cite{cold} and even solids~\cite{solid}.

The Hamiltonian of the quantum Ising model on a general 
lattice (or graph) is given by:
\begin{equation}
H=-\sum_{i,j}  J_{ij}\sigma_i^z \sigma_{j}^z -\sum_{i}h_i\sigma_i^x \label{Hising}
\end{equation}
where $\sigma^k_i$, for $k=x,y,z$, are the Pauli matrices on the state space associated to site 
$i$ (a vertice of the graph). The coefficient $J_{ij}$ represents the strength of interaction 
between sites $i$ and $j$, while $h_i$ represents an external magnetic field applied on site $i$. 
The edges of the lattice determine which systems interact: $J_{ij}\neq 0$ if sites $i$ 
and $j$ are connected by an edge while $J_{ij}=0$ otherwise.

Our main result is a direct proof, for the transverse quantum Ising chain, that if the field in a 
particular site is null, the reduced state of one side of the chain (relative to the site with null 
field) is independent of the Hamiltonian parameters of the other side. So, even if there is a 
strong interaction coupling between each side, their reduced states behaves as if there is none. 

Besides such a proof, we discuss in detail a more physical explanation of the result using 
the duality of the transverse Ising model. The direct proof, however, can be applied to a 
more general setting where the Hamiltonian still has transverse field, but not necessarily at the 
same direction for all sites. Furthermore, it encompasses more general lattices. We 
assume that the lattice can be split in two halves (in the sense that there is no interaction 
between sites of these two halves), and an interface between them (in the the sense that each half 
can interact with the sites of the interface). If the interface has only one site and the field is 
null on it, the result still folds. 

We also investigate whatever the result would still hold if the interface contains more than one site. 
We show an example where the shielding property does not work for positive temperatures, but 
we conjecture that it does work when the system is in the ground state. We show some numerical 
examples 
that corroborate with the conjecture. 

We point out that the dynamics of any many body spin system satisfy similar 
properties, if just a commutation relation is imposed on the Hamiltonian. 

Finally, we discuss some consequences of our results to the effect of local perturbations on both 
the equilibrium and non-equilibrium properties of the perturbed system.

\textit{The shielding property on the Ising chain.}---We consider the Gibbs states at arbitrary 
temperature of the model 
defined by equation 
(\ref{Hising}) on finite open chains. Namely, we take the spins to be embedded on a straight 
line, where they interact only with their first neighbors, so we can use integer numbers $i$ to 
index each site. We assume the couplings and fields to be arbitrary, with the exception that the 
field must be null in some particular site $L$. 
We show that the reduced state of one side, say the sites to the right of site $L$ (those with 
$i\geq 
L$), have no dependence on the parameters of the Hamiltonian of the other side, that is, the sites 
with $i<L$. We will refer to this feature a \emph{shielding 
property}.

\begin{theorem} \label{theo}
Let a chain of $N$ sites be described by the transverse Ising model. Suppose that for some fixed 
site $L$ we have $h_L=0$. If the state $\rho$ of the chain is the Gibbs state, then the reduced 
state of sites $L,...,N$ has no dependence on $h_1,...,h_{L-1},J_1,...,J_{L-1}$, and is given by
\begin{equation}
\rho_{L,...,N}=\frac{e^{-\beta H''}}{\text{\normalfont{Tr}}(e^{-\beta H''})}, \label{eqteo}
\end{equation}
where $H''$ is given by
\begin{align}
H''=-\sum_{i=L} ^{N-1} (J_i\sigma_i^z \sigma_{i+1}^z+h_{i+1}\sigma_{i+1}^x). \label{Hduaslinhas}
\end{align}
defined on the space of sites $L,...,N$.
\end{theorem}

The detailed proof can be found in the Appendix. It is worth 
mentioning, 
however, that it explores the fact that the Hamiltonian can be written as $H=H^I+H^{II}$, where 
 $H^{I}$ and  $H^{II}$ commute and the intersection of their (spatial) supports 
contains only site 
$L$. However, these conditions are not sufficient for the validity of the Theorem. As an 
example, take a chain of $n$ sites with the Hamiltonian 
$H=-h_1\sigma_1^z-\sigma_1^z\sigma_2^z-...-\sigma_{n-1}^z\sigma_n^z$. Suppose that the state of the 
chain is the Gibbs state for some $\beta$. Note that the external magnetic field on almost all sites 
are null, except on site $1$. We can find that the reduced state at site $i$, for $i=1,...,n$, is 
given by
\begin{equation}
\rho_i=\frac{1}{2}\mathds{1}+\frac{1}{2}\text{tanh}^{i-1}(\beta)\text{tanh}(h_1\beta)\sigma_i^z. 
\label{counterr}
\end{equation}
So, the reduced state on every site of the lattice depends on the external magnetic field $h_1$ on 
site $1$ when the system is on some Gibbs state. This illustrates that Theorem~\ref{theo} 
indeed explores the specific structure of the transverse Ising model Hamiltonian.

\textit{Duality on the Ising chain}---The result of 
Theorem~\ref{theo} can also be 
explained, and in a more intuitive way, by the duality of the quantum Ising chain \cite{duality}. 
The duality allows us to write the Hamiltonian in terms of a Hamiltonian for a dual chain where the 
parameters $h$'s and $J$'s swap their roles. To be more precise, define the operators:
\begin{equation}
\mu_{i+1/2}^z=\sigma_i^z\sigma_{i+1}^z,\ \ \ \ 
 \mu_{j+1/2}^x=\prod_{k=j+1}^N\sigma_k^x \label{dualoperators}
\end{equation}
for $i=1,...,N-1$ and $j=1,...,N$. Set $\mu_{N+1/2}^z=\sigma_N^z$, $\mu_{1/2}^z=\sigma_1^z$ and 
$\mu_{1/2}^x=\mathds{1}$. With these definitions, the Hamiltonian (\ref{Hising})  can be 
written as:
\begin{equation}
H=-\sum_{j=1}^{N-1} J_j\mu_{j+1/2}^z-\sum_{j=1}^Nh_j\mu_{j-1/2}^x\mu_{j+1/2}^x. \label{dual}
\end{equation}

Since the operators $\mu^x$ and $\mu^z$ satisfy the algebra of Pauli operators, 
 the Hamiltonian $H$ written as in equation (\ref{dual}) can be seen as the Hamiltonian of a dual 
chain. Since $h_j$ appears multiplying $\mu_{i-1/2}^x\mu_{i+1/2}^x$, it can be interpreted as the 
strength of the interaction between the dual sites, and as $J_j$ appears multiplying 
$\mu_{j+1/2}^z$ only, it can be interpreted as the external magnetic field. 

\begin{figure}[h]
\includegraphics[scale=0.7]{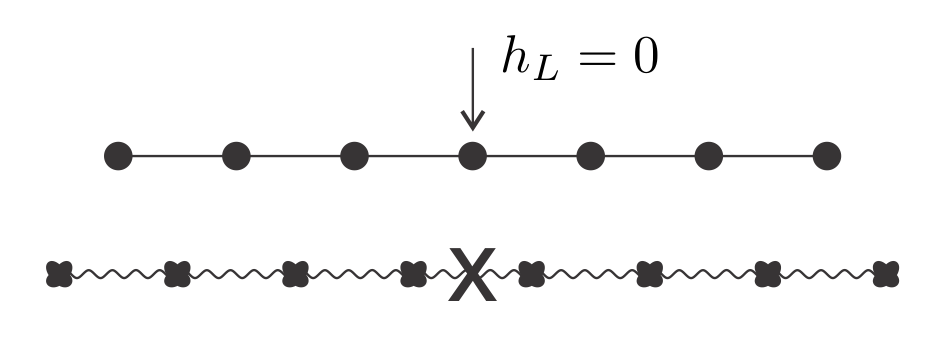}
\caption{Representations of the original and of the dual chains. The sites of the dual chain 
correspond to the links of the original chain and vice-versa.} \label{dualchainmn}
\end{figure}

Therefore, as we assume $h_L=0$, for some $L=2,...,N-1$, the dual chain has two decoupled halves 
(see figure \ref{dualchainmn}). We can then safely conclude that the reduced states of each side of 
the dual chain do not depend on the parameters of the Hamiltonian of the other side, since the 
whole 
state is a product of the Gibbs states of each half.

To arrive on our desired conclusions for the original chain, we can explore the fact that all the 
local observables of one side of the original chain is a combination of observables of only that 
side of the dual chain. This is possible to show by finding the inverse of 
Equations~(\ref{dualoperators}). Since the reduced state of the dual chain is a product state, a 
local observable of one side of the original chain can be written only as a function of the 
parameters of the Hamiltonian on the same side. We conclude that the reduced state 
$\rho_{1,...,L-1}$ of the original chain does not depend on the parameters 
$h_L,....,h_{N},J_L,...J_{N}$. This argument, however, does not clearly show the desired property 
for the reduced state containing site $L$, although Theorem \ref{theo} ensures it must also hold in 
that case.

\begin{figure}[h]
\includegraphics[scale=0.5]{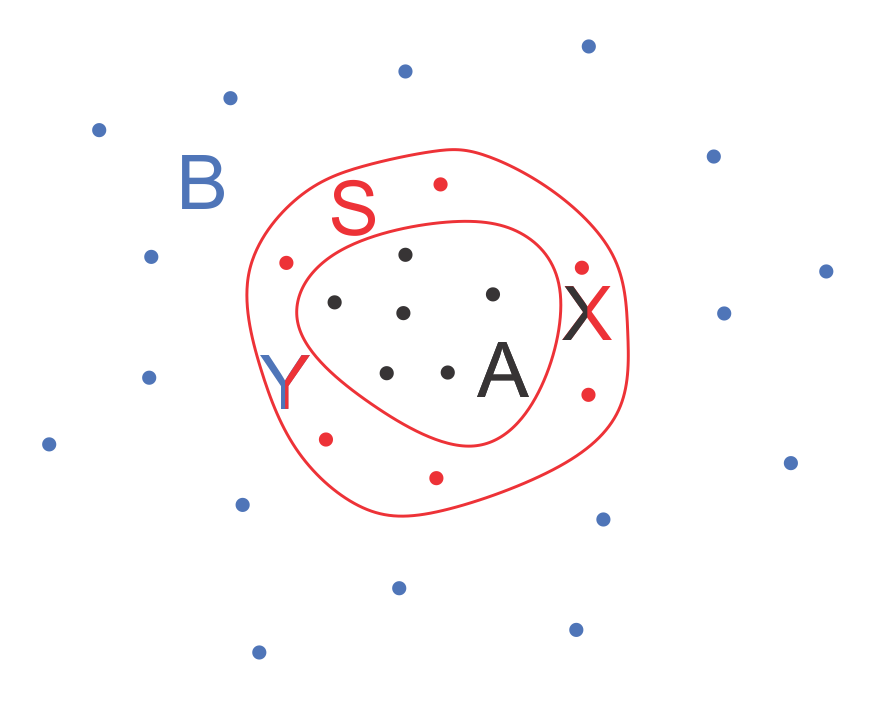} \includegraphics[scale=0.7]{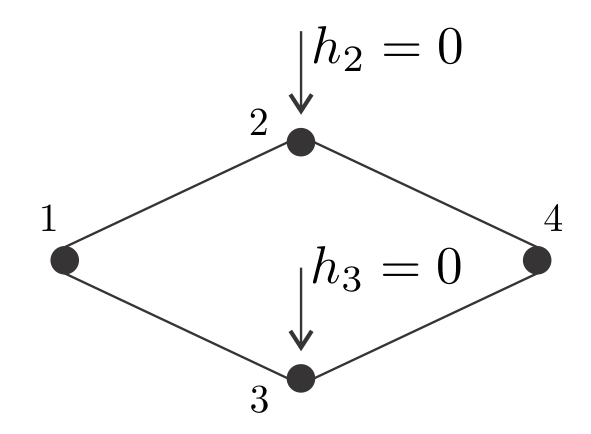}

 a)   \hspace{3.5cm}  b)
\caption{a) An example of sets $A$, $X$, $S$, $B$ and $Y$. b) Example of a system
with two sites on the interface which does not satisfy the shielding property. 
The reduced state on site $4$ has dependence on the external magnetic field $h_1$ applied on site 
$1$.} \label{conj} 
\end{figure}

\textit{The shielding property on general lattices}---The proof of Theorem \ref{theo} can be 
generalized to more 
general lattices and slightly more general Hamiltonians. Let $\Lambda$ be a 
lattice that can be 
spit into two sets $X$ and $Y$ where $X\cup 
Y=\Lambda$ and $X\cap Y=S$. Furthermore, assume that all sites $i\in X$ 
and $j\in Y$, such that $i,j\notin S$, are not connected by an edge. If $A=X-S$ and $B=Y-S$, we 
call $S$ the \emph{interface} between $A$ and $B$, or between $X$ and $Y$. See Fig.~\ref{conj} for 
a schematic representation of all 
these sets. We have then: 

\begin{theorem} \label{theo'}
Let $\Lambda$ be a lattice as described above where, furthermore, $S=\{L\}$ for some site $L$. 
Assume the system 
Hamiltonian is:
\begin{equation}
H=-\sum_{i,j}  J_{ij}\sigma_i^z \sigma_{j}^z 
-\sum_{i}h_i\sigma_i^x-\sum_{i}g_i\sigma_i^y, \label{eqHxy}
\end{equation}
where $h_L=g_L=0$. For any temperature, the reduced state on the set $Y$ of the Gibbs state of 
the whole lattice has no 
dependence on $h_i$, $g_i$ and $J_{i,j}$, for all $i,j\in X$. Furthermore, the reduced state is 
given by
\begin{equation}
\rho_{Y}=\frac{e^{-\beta H''}}{\text{\normalfont{Tr}}(e^{-\beta H''})},
\end{equation}
where $H''=-\sum_{i,j\in B} (J_{ij}\sigma_i^z \sigma_{j}^z +h_{j}\sigma_{i}^{x}+g_{i}\sigma_{i}^y)$.
\end{theorem}

This shows that if a system is described by the transverse quantum Ising model on a lattice 
which separates two regions by only one site, the shielding property is satisfied. Note that in 
each side of the lattice we can even have long range interactions between sites. Moreover, the 
transverse field may vary from site to site, as long as it is always transverse to the interaction 
direction. 

One could wonder if this shielding effect occurs when the interface contains more than one site. 
In general, it is not the case. Consider a system with four 
sites, as depicted in Figure~\ref{conj}b), with Hamiltonian 
$$H=-\sigma_1^z\sigma_2^z-\sigma_2^z\sigma_3^z-h_1\sigma_1^x-\sigma_2^z\sigma_4^z-\sigma_3^z\sigma_4
^z-h_4\sigma_4^x.$$

We can take $X=\{1,2,3\}$ and $Y=\{2,3,4\}$, with the interface given by 
sites $2$ and $3$. If the state of the system is given by the Gibbs state with $\beta>0$, we have shown that the expected value of the 
magnetization $\langle\sigma_4^x\rangle$ on site $4$ has a dependence on the external magnetic field 
$h_1$, applied on site $1$ (see the Appendix for further details). However, if we take $\beta$ tending to infinity, 
$\langle\sigma_4^x\rangle$ is independent of $h_1$. That is, it seems that the property 
is still valid when the system is in the ground state.

\begin{figure}[h]
\includegraphics[scale=0.25]{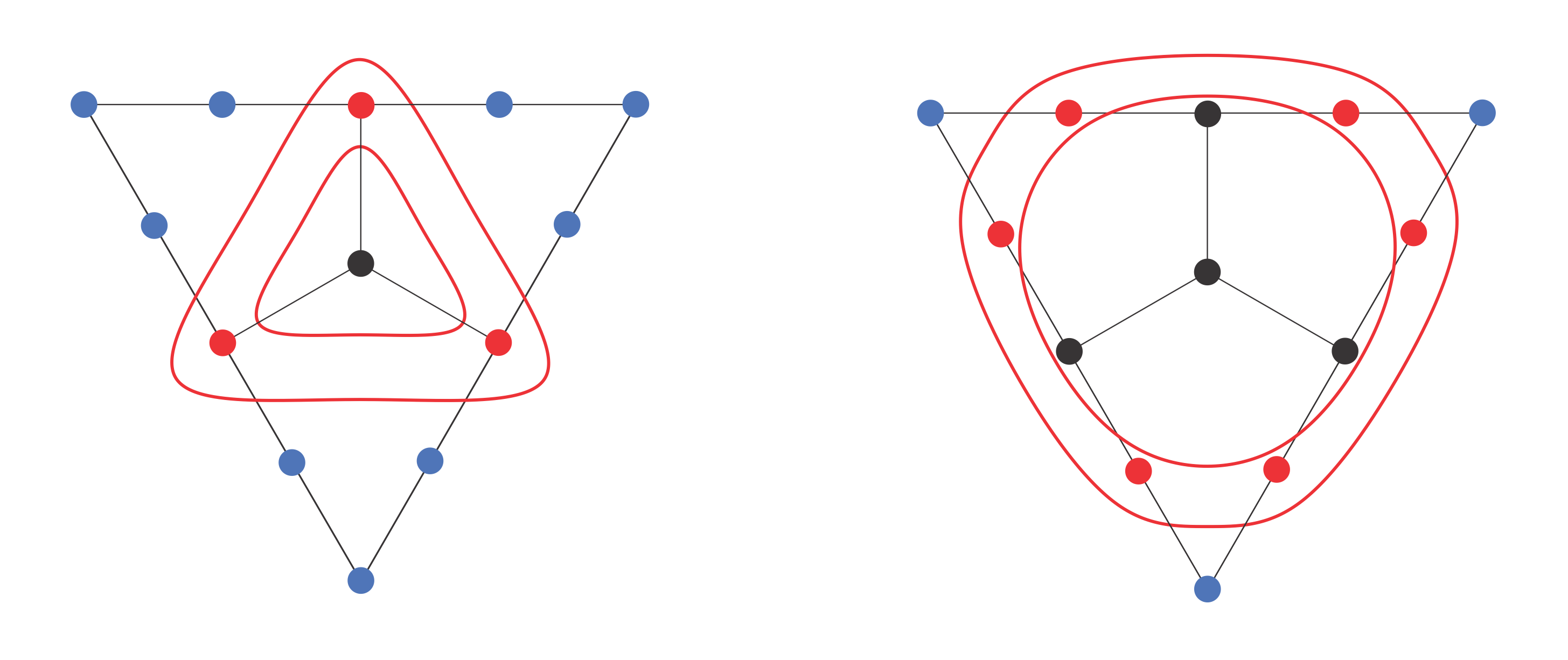}
\caption{Two different arrangements are considered here for the same lattice. The subset between 
the red lines is the set $S$, where the external magnetic field is null. Outside the red lines there 
is the subset $B$ and inside the set $A$.} \label{triang}
\end{figure}

We have considered two additional examples, as shown on Figure~\ref{triang}, to explore if 
the shielding property would still hold, however, for ground states. For each 
arrangement the lattice is 
the same, but the sets $A$, $B$ and $S$ are different. For each of them, the external magnetic 
field applied on the sites of $S$ is null. The interaction parameters of the 
Hamiltonian~\eqref{Hising} were chosen as $J_{ij}=1$ for every connected pairs $i,j$. 

By exact numerical diagonalization of the Hamiltonian, we have seen that even if we modify the 
external magnetic field on set $B$, the magnetization of 
each site in $A$, for the ground state of the system, apparently remains the same. 
We have constructed 
several Hamiltonians where, first, we set the field in sites of $A$ 
randomly between $0$ and $1$ (the values of the fields were drawn independently from the uniform 
distribution on the interval $[0,1]$). Then we constructed several distinct Hamiltonians by 
randomly selecting the field in sites of $B$ (independently and according to the uniform 
distribution on $[0,1]$) plus an homogeneous field in the whole region. In each instance we have 
calculated the ground state via exact diagonalization and we computed the expected value of the 
magnetizations of all sites of $A$. Within numerical accuracy, no variation of the magnetization of 
sites of $A$ were detected with the variation of the parameters on $B$. 

These examples show that it is reasonable to believe that, for ground states specifically, the 
shielding property works for systems 
which the interface contain more than one site.
Then, we state the following:

\begin{conj}
Let $\Lambda$ be a lattice that can be divided into two sets $X$ and $Y$ such that $X\cup 
Y=\Lambda$ and $X\cap Y=S$. Furthermore, assume the sites $i\in X$ and $j\in Y$ such that 
$i,j\notin S$ are not connected. Suppose there is a system which can be 
described by this lattice with the Hamiltonian
\begin{equation}
H=-\sum_{i,j}  J_{ij}\sigma_i^z \sigma_{j}^z 
-\sum_{i}h_i\sigma_i^x-\sum_{i}g_i\sigma_i^y, \label{eqHxyy}
\end{equation}
and suppose that on the sites $l\in S$ we have $h_l=g_l=0$. If the state of the lattice is the 
ground state, then the reduced state of the set $Y$ has no dependence on $h_i$, $g_i$ and $J_{i,j}$, 
for all $i,j\in X$.\end{conj}

\textit{Dynamics}---As we have seen, the shielding property is true for the Gibbs state of a 
transverse Ising model when the interface between two regions contains only one site. We have also 
seen that the commutation of the Hamiltonian terms corresponding to each region is an important 
feature in the proof. However, it is not a sufficient condition, as we have shown in 
example (see Eq.~\eqref{counterr}). On the other hand, this commutation relation is, in some sense, 
a sufficient condition for an analogous property of the system \emph{dynamics}.

As before, let $X$ and $Y$ be two regions of the lattice, such that 
$X\cup Y=\Lambda$. Let $H$ be \emph{any} many-body Hamiltonian that can be written as $H=H_X+H_Y$, 
where $\text{supp}(H_X)=X$ and $\text{supp}(H_Y)=Y$, and suppose that $[H_X,H_Y]=0$. If 
$\mathcal{O}$ is any observable with $\text{supp}(\mathcal{O})=A=Y-S$, then $H_X$ 
and $\mathcal{O}$ commute. Therefore, the 
expected value of $\mathcal{O}$ at time $t$ is given by:
\begin{align}
\langle \mathcal{O}(t) \rangle_\rho &=\text{Tr}(\rho(t)\mathcal{O}) \nonumber \\ &=\text{Tr}(e^{-it 
H_X}e^{-it H_Y}\rho e^{it H_Y}e^{it H_X}\mathcal{O}) \nonumber \\ 
&=\text{Tr}(e^{-it H_Y}\rho e^{it H_Y}\mathcal{O}). \label{eq1} 
\end{align}
where $\rho$ is any initial state for the system, and we have used on the third equality the cyclic 
property of the trace. We note that the expected value of $\mathcal{O}$ depends only on 
$\mathcal{H}_{Y}$. Since $\mathcal{O}$ is an arbitrary observable of region $A$, it holds that the 
reduced state of the system in that region depends only on the parameters of region $Y$ (assuming 
the initial state does not hold any dependence on the Hamiltonian).

\textit{Discussion}---Our main results, together with the discussion in the previous section, have 
strong implications on the effect of local perturbations on the equilibrium state of systems 
described by the Ising model, as well on its dynamics. 

Consider a local perturbation $W$ on a many-body Hamiltonian $H$. That is, $W$ may be large in 
norm, as long as it has small spatial support. One can show~\cite{roeck,sven}, for instance, that 
the reduced ground states of $H$ and $H+W$ are exponentially similar away from the support of 
$W$~\footnote{That is, the norm of the difference between the perturbed and unperturbed reduced 
states 
of the ground states exponentially decays with the distance between the region where the 
perturbation takes place and the region where we take the reduced state.}. In the setting of 
Theorem~\ref{theo'}, however, we see an extreme behavior for all equilibrium states at all 
temperatures: whatever the perturbation that is done on the parameters of side $X$ of the lattice 
(\emph{i.e.}, any perturbation $W$ that keeps the form of the Hamiltonian), the equilibrium 
state on side $Y$ will remain completely unperturbed. That is, the equilibrium reduced states on 
$Y$ are exactly the same for $H$ and $H+W$.

One can also consider the dynamical effect of the local perturbation. That is, instead of comparing 
the equilibrium states of perturbed and unperturbed Hamiltonian, we may consider the situation 
where the unperturbed system is in equilibrium and, at time $t=0$, its Hamiltonian is instantly 
changed to $H+W$ (that is, a local quantum quench is applied). It is well known, due to 
Lieb-Robinson 
bounds~\cite{LR}, that such perturbations in many body systems with short ranged interactions 
propagate effectively with a finite velocity. In Refs.~\cite{lightcone,Bravyi} this is used to show 
that no (significant) amount 
of information can be transferred from regions $A$ to $B$ of a system by applying local quantum 
quenches in one of the regions, in a interval of time small compared to the distance 
between the regions (scaled by the group velocity of perturbations). In the setting described 
in the last section, due to Eqs.~\eqref{eq1}, we have a much 
stronger effect: no information whatsoever can be sent between regions $A$ and $B$, no matter how 
much time is available to the process.

\begin{figure}[h]
\includegraphics[scale=0.3]{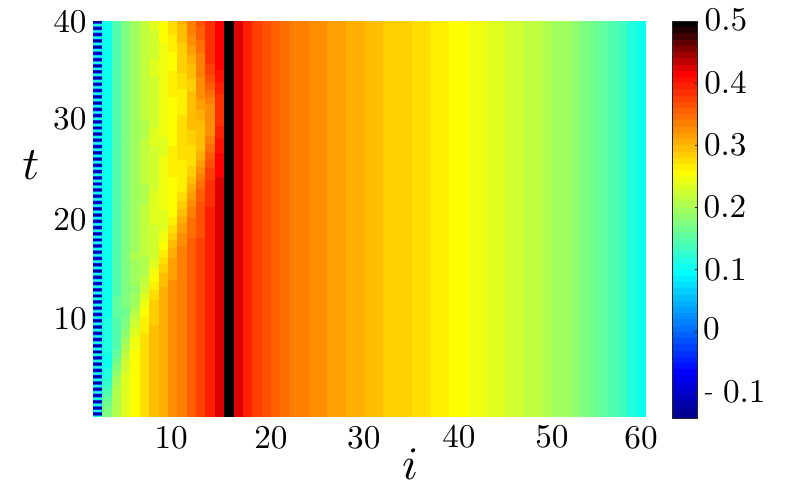}
\caption{Temporal evolution of the magnetization of each site of a transverse quantum Ising chain. 
The external magnetic field of site $15$ was fixed null and the evolution was calculated via tDMRG.} 
\label{din}
\end{figure}

As a visual illustration of this effect, we have simulated, via $t-$DMRG, the evolution of a 
transverse Ising model on a chain with 60 sites with initial parameters $J_i=1$ for all the sites, 
$h_{15}=0$ and $h_i=\frac{1}{2}$ for $i\neq 15$. The system is initially prepared in the ground 
state, when we perform a local quench in the first site, changing its magnetic field to $h_1=-10$. 
We show in Fig.~\ref{din} the magnetization of each site of the chain as a function of time. At the 
left part of the plot we see the perturbation propagating and 
being reflected on site $15$, where we have made the external magnetic field null. On the right 
part we can see that the magnetization of all the sites after site $15$ remain unaltered.

\textit{Conclusion.}---We have defined and shown the shielding property for the transverse Ising 
model. We provide a direct proof of it, which is valid for 
arbitrary parameters of the Hamiltonian in each side of the lattice. For the special case of a 
chain, we further explain the property using the duality of the model.

When the interface between the two halves of the lattice has more than one site we show 
an example where the shielding property does not work for positive temperatures. It seems, 
however, that it still works at null temperature. Finally,  we have explored the consequences of 
such results if the parameters of the 
Hamiltonian are perturbed in one side of the lattice. We show that, no matter how significant this 
perturbation is, the other side of the lattice is unchanged, both in its equilibrium state, as in 
its dynamics. 

\textit{Acknowledgements.}---We acknowledge financial support from Conselho Nacional de Desenvolvimento Cient\'ifico e Tecnol\'ogico (CNPq) and Coordena\c{c}\~ao de Aperfei\c{c}oamento de Pessoal de N\'ivel Superior (CAPES). We thank Rodrigo G. Pereira, Diego B. Ferreira and Sheilla de Oliveira M. for useful discussions.

\appendix
\section{Proof of Theorem \ref{theo}}

We first rewrite the Hamiltonian in the following way. If on Equation (\ref{Hising}) 
we 
have that $h_L=0$ for some $L=2,...,N-1$, we can set $H=H^I+H^{II}$, where
\begin{equation}
H^I=-\sum_{i=1} ^{L-1} (J_i\sigma_i^z \sigma_{i+1}^z +h_i\sigma_i^x) \label{Hlinha}
\end{equation}
and
\begin{equation}
H^{II}=-\sum_{i=L} ^{N-1} (J_i\sigma_i^z \sigma_{i+1}^z +h_{i+1}\sigma_{i+1}^x). \label{Hlinhas}
\end{equation}

Note that $H^I$ has support on the space of sites $1,...,L$, while $H^{II}$ has support on the 
space of $L,...,N$. However, even though they have intersecting supports, they
commute. In any case, we can write:
\begin{equation}
H^I= H'\mathds{1}_{L+1}...\mathds{1}_N \label{hlinha}
\end{equation}
and
\begin{equation}
H^{II}=\mathds{1}_1...\mathds{1}_{L-1} H'' \label{hlinhas}
\end{equation}
where $H'$ is defined on the space of sites $1,...,L$ while $H''$ on the space of $L,...,N$, and we 
are omitting the tensor product among operators. 

With all of the above in mind, we can write then
\begin{align}
&e^{-\beta H}=e^{-\beta H^I}\cdot e^{-\beta H^{II}} \\
=(e^{-\beta H'}&\mathds{1}_{L+1}...\mathds{1}_N)\cdot(\mathds{1}_1...\mathds{1}_{L-1} e^{-\beta 
H''}). \label{equationgibbs}
\end{align}
Therefore, the reduced density operator of the Gibbs state is, on sites $L,...,N$, satisfy:
\begin{align}
\rho_{L,...,N}&\propto\text{Tr}_{1,...,L-1}(e^{-\beta H}) \\ 
=\{\text{Tr}_{1,...,L-1}(e&^{-\beta H'})\mathds{1}_{L+1}...\mathds{1}_N)\}\cdot e^{-\beta H''}. 
\label{partialtrace}
\end{align}
We will show that the partial trace of $e^{-\beta H'}$, appearing on Eq.~\eqref{partialtrace}, is a 
multiple of the identity $\mathds{1}_L$, so the theorem follows after normalization. Writing the 
exponential $e^{-\beta H'}$ as its series expansion, the partial trace becomes
\begin{equation}
\text{Tr}_{1,...,L-1}(e^{-\beta H'})=\sum_{n=0}^\infty \frac{\beta^n}{n!}\{\text{Tr}_{1,...,L-1}(H'^n)\}. \label{eqtraco}
\end{equation}
To reach the desired result it suffices to show that $\text{Tr}_{1,...,L-1}(H'^n)=c_n\mathds{1}_L$, 
for all $n=0,1,2,...$, where $c_n$ is some constant. The case $n=0$ is trivial, so let us 
take care of positive values of $n$. First, we have that
\begin{align}
H'^n=\Bigg(-\sum_{i=1} ^{L-1} (J_i\mathds{1}_1...\mathds{1}_{i-1}\sigma_i^z 
\sigma_{i+1}^z\mathds{1}_{i+2}...\mathds{1}_{L} \nonumber \\ 
+h_i\mathds{1}_1...\mathds{1}_{i-1}\sigma_i^x\mathds{1}_{i+1}...\mathds{1}_{L})\Bigg)^n. \label{H'n}
\end{align}
Note that $H'=\tilde{H}_{(1)}\mathds{1}_L+\bar{H}_{(1)}\sigma_L^z$, where $\tilde{H}_{(1)}$ and 
$\bar{H}_{(1)}$ are defined on the space of the sites $1,...,L-1$. So any power of $H'$ will have 
this form, that is:
\begin{equation}
H'^n=\tilde{H}_{(n)}\mathds{1}_L+\bar{H}_{(n)}\sigma^z_L.\label{eqHn}
\end{equation}
More explicitly, we have that
\begin{align}
H'^n&=\left(\sum_{{\eta_1},...,\eta_{L-1}} 
a_{(n,{\eta_1},...,\eta_{L-1},\mathds{1}_{L})}\eta_1\eta_2...\eta_{L-1}\right)\mathds{1}_L \nonumber 
\\ &+\left(\sum_{{\eta_1},...,\eta_{L-1}} 
a_{(n,{\eta_1},...,{\eta_{L-1}},{\sigma^z})}\eta_1\eta_2...\eta_{L-1}\right)\sigma^z_L \nonumber 
\\ \ \nonumber
\end{align}
where the sum ranges over variables $\eta_1,...,\eta_{L-1}$ and each of these $\eta_i$ assumes 
the values $\sigma^z_i, \sigma^x_i, \sigma^y_i$ and $\mathds{1}_i$.
Writing in a more concise way:
\begin{equation}
H'^n=\sum_{\eta_1,...,\eta_L} a_{(n,{\eta_1},...,\eta_{L})}\eta_1\eta_2...\eta_L, \label{eqh}
\end{equation}
where the sum is made on the variables $\eta_1,...,\eta_{L}$. For $i=1,...,L-1$, the variable 
$\eta_i$ assumes the values $\sigma^z_i, \sigma^x_i, \sigma^y_i$ and $\mathds{1}_i$, and the 
variable $\eta_L$ assumes only the values $\sigma^z_L$ and $\mathds{1}_L$.

We will prove that each term of $\bar{H}_n$ has null trace. For concreteness and ease of notation, 
we will detail the argument for the term $\bar{H}_{(3)}$, where the field is null at site 
$L=3$ and the chain has $J_{1}=J_{2}=1$. The proof for any term of $H'^n$, for all $n=1,2...$, 
all values of $L$ (between $2$ and $N-1$), will follow naturally the same steps. 

With these assumptions, Eq.~\eqref{eqHn} becomes
\begin{align}
H'^3&=-(\sigma_1^z\sigma_2^z+h_1\sigma_1^x+\sigma_2^z\sigma_3^z+h_2\sigma_2^x)^3\nonumber \\
&=-\{(4+h_1^2+h_2^2)\sigma_1^z\sigma_2^z+ (4+h_1^2+h_2^2)h_1\sigma_1^x\mathds{1}_2 \nonumber \\
&+(2+h_1^2+h_2^2)h_2 \mathds{1}_1\sigma_2^x
-2h_1h_2\sigma_1^y\sigma_2^y \} \mathds{1}_3 \nonumber \\ &-\{(2+h_1^2+h_2^2)\sigma_1^z\mathds{1}_2+
(2+2h_1^2)\mathds{1}_1\sigma_2^z\nonumber \\
&-2ih_1h_2\sigma_1^x\sigma_2^y+2h_2\sigma_1^z\sigma_2^x\}\sigma_3^z. \label{eqex}
\end{align}
Here, $\bar{H}_{(3)}$ is the operator in the curly brackets of the 
last two lines of Eq.~\eqref{eqex}. We can see that its trace is null due to the fact that every 
one of its terms has at least one Pauli matrix in its factors. We will show why this must be the 
case.

\begin{figure}[h]
\includegraphics[scale=1]{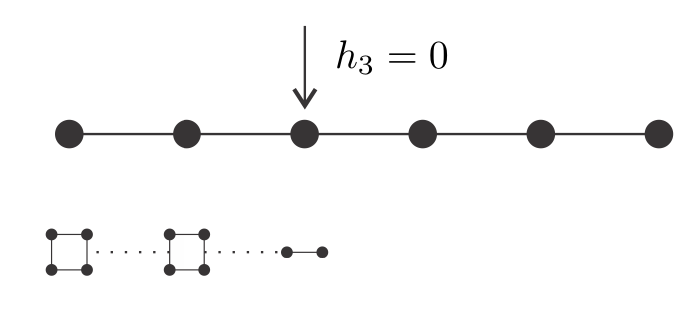}
\caption{Board game associated to a chain when $h_3=0$.}\label{gamechain}
\end{figure}

\begin{figure}[h]
\includegraphics[scale=0.7]{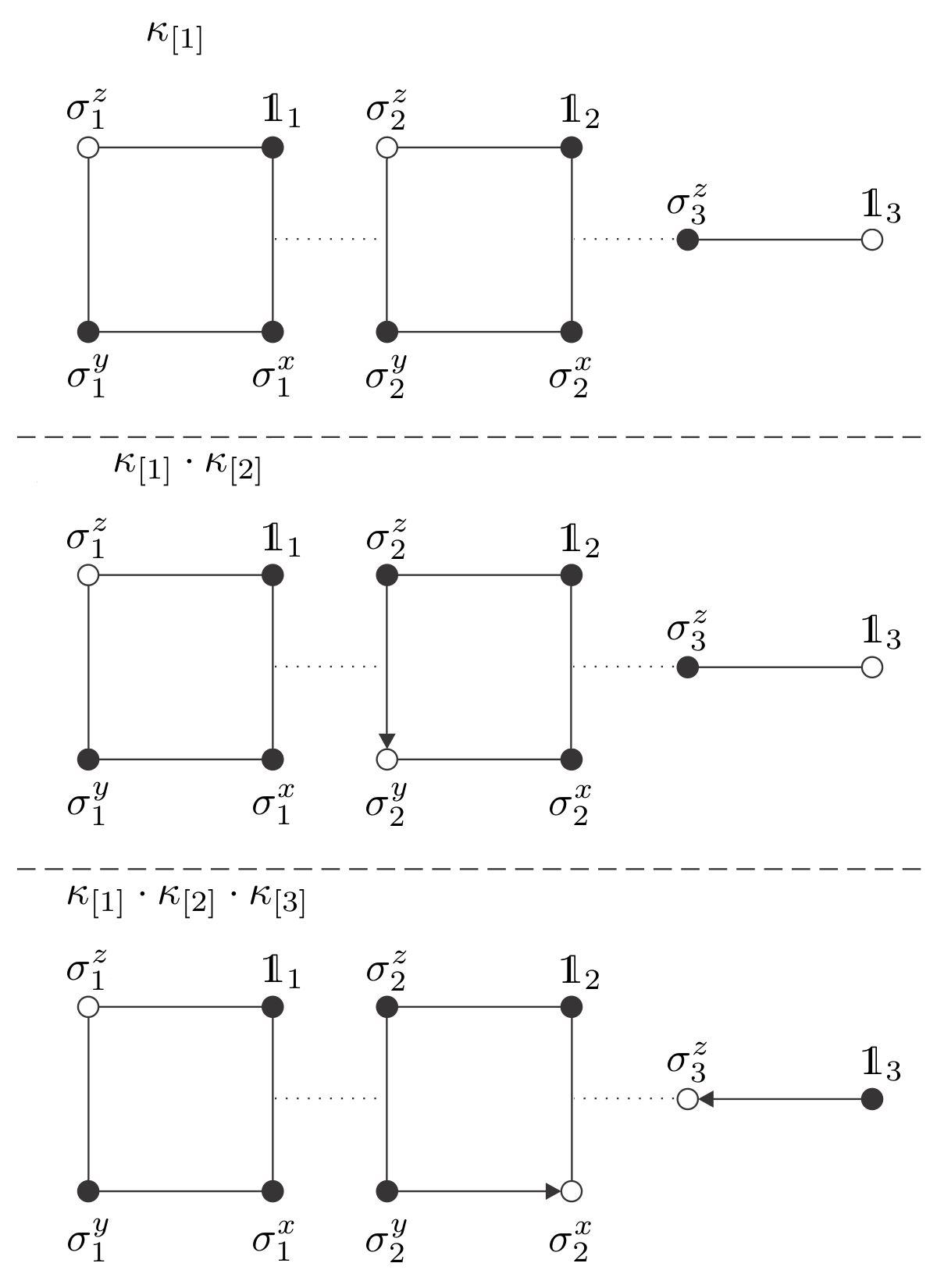}
\caption{Scheme to understand the product $\kappa_{[1]}\kappa_{[2]}\kappa_{[3]}$ as a ``board 
game''. The rows represent different steps ($n=3$) of a board game with $L=3$ white pieces. The 
pieces can move from one vertex to another only when they are connected by continuous lines. 
}\label{tab}
\end{figure}

Take for instance the last term of Eq.~\eqref{eqex}, $\sigma^z_1\sigma^x_2\sigma^z_3$. It was 
obtained by the multiplication of the terms $\sigma_1^z\sigma_2^z\mathds{1}_3$, 
$\mathds{1}_1\sigma_2^x\mathds{1}_3$ and $\mathds{1}_1\sigma_2^z\sigma_3^z$, which we will call 
$\kappa_{[1]}$, $\kappa_{[2]}$ and $\kappa_{[3]}$. That is:
\begin{align}
\kappa_{[1]}\cdot\kappa_{[2]}\cdot\kappa_{[3]}&=(\sigma^z_1\sigma^z_2\mathds{1}_3)\cdot
(\mathds{1}_1\sigma^x_2\mathds{1}_3)\cdot(\mathds{1}_1\sigma^z_2\sigma^z_3) \nonumber \\
&\propto (\sigma^z_1\sigma^y_2\mathds{1}_3)\cdot(\mathds{1}_1\sigma^z_2\sigma^z_3)\nonumber \\
&\propto (\sigma^z_1\sigma^x_2\sigma^z_3)=\eta_1\eta_2\eta_3. \label{eqtab}
\end{align}
Note that the products of $\eta$'s are tensorial products, while the products of $\kappa$'s are 
matrix products. We can represent schematically, as in Figures~\ref{gamechain} and \ref{tab}, the 
product in Eq.~\eqref{eqtab}, as a ``board game''. 

In Figure \ref{gamechain} we show how the game is constructed. For sites $1$ and $2$ we associate a 
square, while for site $3$, where $h_3=0$ we associate a line. We put $L=3$ white pieces on 
this game, occupying the vertices of the squares and the lines. We will label these vertices as 
$\sigma_i^x$, $\sigma_i^y$, $\sigma_i^z$ and $\mathds{1}_i$, for $i=1,2,3$.

In figure \ref{tab} we show example steps of the game, where each row is one step. We put one white 
piece for each site $i$, for $i=1,2,3$, and these pieces can move on the square or line associated 
to this site. In the first row we put these pieces positioned on the labelled spaces in a way that 
the term $\kappa_{[1]}=\sigma_1^z\sigma_2^z\mathds{1}_3$ is represented.
In the next step (second row) we change the positions of these pieces to represent 
$\kappa_{[1]}\cdot\kappa_{[2]}$ which is proportional to $\sigma_1^z\sigma_2^y\mathds{1}_3$. And 
again (third row) we change their positions to represent 
$\kappa_{[1]}\cdot\kappa_{[2]}\cdot\kappa_{[3]}$ which is proportional to 
$\sigma_1^z\sigma_2^x\sigma_3^z$.

We will say that a piece is on the left when it is in some position labeled by $\sigma_i^z$ or 
$\sigma_i^y$. At the beginning we start with two pieces on the left, an even number. Multiplying 
$\kappa_{[1]}$ by $\kappa_{[2]}$, we move one piece vertically, without changing the number 
of pieces on the left side. Now, multiplying $\kappa_{[1]}\cdot\kappa_{[2]}$ by $\kappa_{[3]}$ we 
change the side of two pieces, so the number of pieces on the left is still even. As we have an even 
number of pieces on the left, at least one of the pieces corresponding to sites $1$ or $2$ have to 
be different of identity, and this guarantees that the partial trace is null.

We can easily generalize the argument for the general case. Fig.~\ref{tab} would be similar, 
but with $L$ pieces and $n$ rows. The important fact is that: it does not matter the positions of 
the pieces corresponding to $\kappa_{[1]}\cdot...\cdot\kappa_{[l-1]}$, if 
$\kappa_{[l]}=\sigma_i^x$, it will just move the $i$-th piece vertically, and if 
$\kappa_{[l]}=\sigma_i^z\sigma_{i+1}^z$ it will just change the side of pieces $i$ and $i+1$. 
Furthermore, these are the only possible ``moves''. As $\kappa_{[1]}$ always have an even number of 
pieces on the left and the allowed moves just change the side of an even number of pieces, we have 
that the product $\kappa_{[1]}...\kappa_{[L]}$ (proportional to $\eta_1...\eta_L$) just have an even 
number of pieces on the left. Then, if $\eta_L=\sigma_L^z$, at least one $\eta_i$, for 
$i=1,...,L-1$, is a Pauli matrix, and then 
\begin{align}
&\text{Tr}_{1,...,L-1}(\eta_1...\eta_{L-1}\sigma_L^z) \nonumber \\
&=\text{Tr}(\eta_1...\eta_{L-1})\sigma_L^z=0
\end{align}
which implies that $\text{Tr}(\bar{H}_{(n)})=0$. Therefore, by equation (\ref{eqHn}) we have that 
$\text{Tr}_{1,...,L}(H'^n)=\text{Tr}(\tilde{H}_{(n)})\mathds{1}_L$, which implies that 
$\text{Tr}_{1,...,L}(e^{-\beta H'})=c\mathds{1}_L$. \qed

\section{Generalization of Theorem 1} \label{teo1'}

Here we discuss the proof of Theorem~\ref{theo'}. It generalizes Theorem~\ref{theo} for more 
general lattices (illustrated in Fig.~\ref{gen}), and, for the sake of completeness, 
we restate it:

\begin{figure}[h]
\includegraphics[scale=0.65]{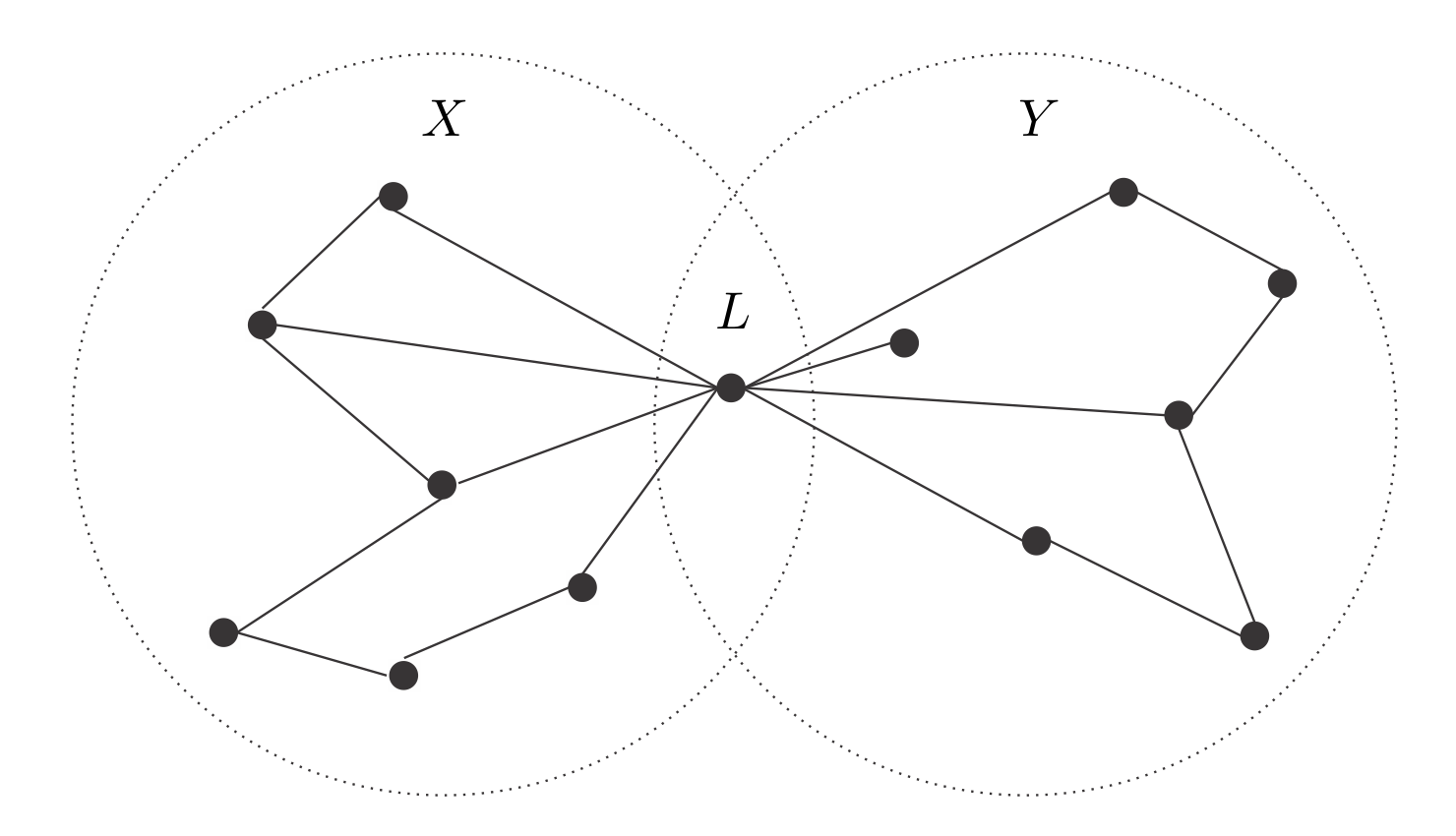}
\caption{An example of a more general lattice}\label{gen}
\end{figure}
\begin{figure}[h]
\includegraphics[scale=0.65]{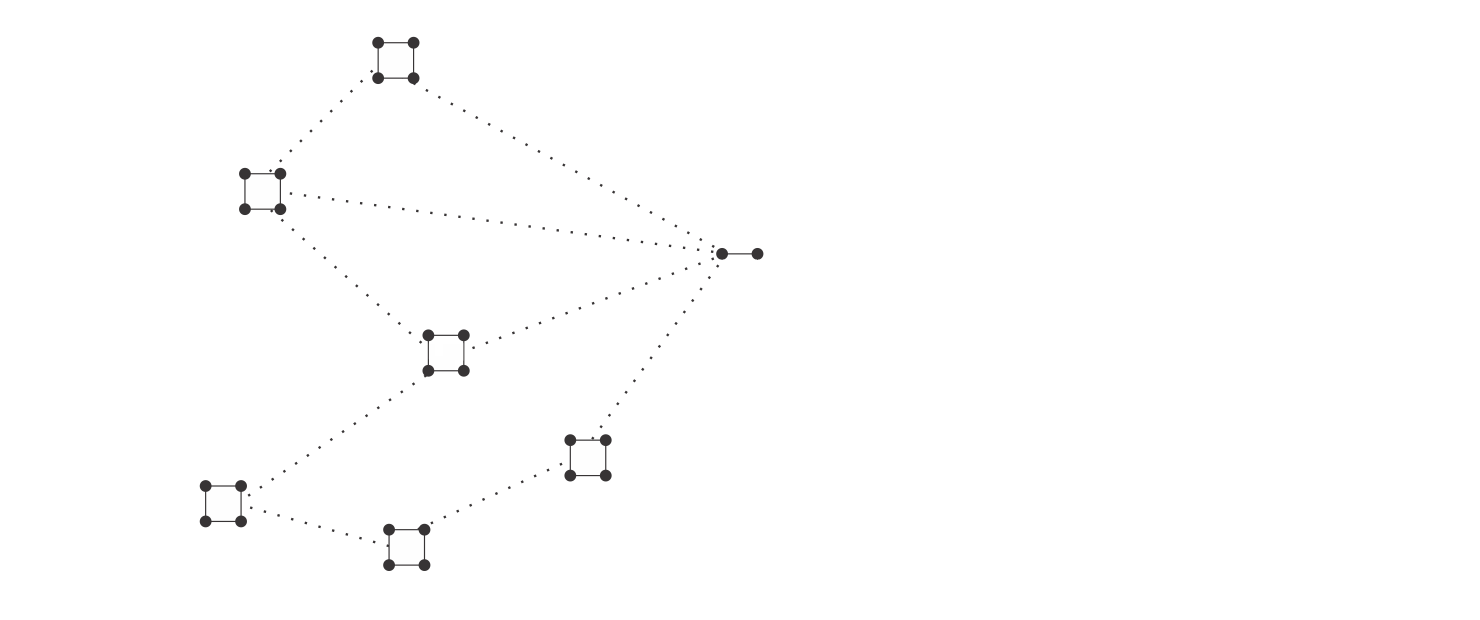}
\caption{Scheme for the proof analogous as in figure \ref{tab}.}\label{geng}
\end{figure}

\renewcommand{\thetheorem}{2'}
\begin{theorem} \label{theo'}
Let a lattice $\Lambda$, which can be divided into two sets $X$ and $Y$ such that $X\cup Y=\Lambda$ 
and $X\cap Y=\{L\}$, where $L$ denotes a single site. Furthermore, the sites $i\in X$ and $j\in Y$ 
such that $i,j\neq L$ are not connected to each each other. Suppose there is a system which 
can be described by this lattice with the Hamiltonian
\begin{equation}
H=-\sum_{\langle i,j\rangle }  J_{ij}\sigma_i^z \sigma_{j}^z -\sum_{i}h_i\sigma_i^x-\sum_{i}g_i\sigma_i^y. \label{eqHxy}
\end{equation}
and suppose that on the site $L$ we have $h_L=g_L=0$. If the state of the lattice is the Gibbs 
state, then the reduced state of the set $Y$ has no dependence on $h_i$, $g_i$ and $J_{i,j}$, for 
all $i,j\in X$. Furthermore, this reduced state is given by
\begin{equation}
\rho_{Y}=\frac{e^{-\beta H''}}{\text{Tr}(e^-\beta H'')},
\end{equation}
where $H''=-\sum_{i,j\in Y} (J_{ij}\sigma_i^z \sigma_{j}^z +h_{j}\sigma_{i}^{x}+g_{i}\sigma_{i}^y)$.
\end{theorem}

\emph{Proof:} The proof is along the same lines of Theorem \ref{theo}. When the lattice is 
not a chain, the argument of the board game is analogous to the argument for chains and it is 
pictured on figure~\ref{geng}. The fact that we have an even number of pieces on the left does not 
depend on the relative geometry of the squares, so this fact is true for these alternative board 
game and the previous proof follows naturally.

Now, the Hamiltonian of this theorem is apparently a bit different from the Hamiltonian of the 
first theorem, because of the terms involving $\sigma_i^y$. But, note that 
\begin{align}
h_{j}\sigma_{i}^{x}+g_{i}\sigma_{i}^y&=\sqrt{h_i^2+g_i^2}\times  \nonumber \\ 
&\Bigg(\frac{h_i}{\sqrt{h_i^2+g_i^2}}\sigma_{i}^{x}+\frac{g_i}{\sqrt{h_i^2+g_i^2}}\sigma_{i}^{y}
\Bigg) \nonumber \\
&=\sqrt{h_i^2+g_i^2}\sigma_i^{a_ix+b_iy}=\sqrt{h_i^2+g_i^2}\sigma_i^{x_i}, \nonumber
\end{align}
where $a_i=h_i/\sqrt{h_i^2+g_i^2}$ and $b_i=g_i/\sqrt{h_i^2+g_i^2}$ and $x_i=a_ix+b_iy$ is some 
direction perpendicular to $z$. So the Hamiltonian (\ref{eqHxy}) can be written as
\begin{equation}
H=-\sum_{\langle i,j\rangle }  J_{ij}\sigma_i^z \sigma_{j}^z -\sum_{i}\sqrt{h_i^2+g_i^2}\sigma_i^{x_i}
\end{equation}
Define $y_i$ such that $\sigma^z\sigma^{x_i}=\sigma^{y_i}$. We have that the operators 
$\sigma^z,\sigma^{x_i}, \sigma^{y_i}$ and $\mathds{1}$ satisfy the same algebra as 
$\sigma^z,\sigma^{x}, \sigma^{y}$ and  $\mathds{1}$. Furthermore these new operators associated to 
the space of different sites still commute between each other, that is, $[\sigma_i^a,\sigma_j^b]=0$ 
if $i\neq j$, for $a=x_i,y_i,z$ and $b=x_j,y_j,z$. Then all the arguments of the previous proof are 
valid here too.
\qed
 
\section{Calculations of the Lattice with Two Sites in the Interface}

In this section we exhibit the calculations of the example shown in figure \ref{conj}b. Its Hamiltonian is given by 
$$H=-\sigma_1^z\sigma_2^z-\sigma_2^z\sigma_3^z-h_1\sigma_1^x-\sigma_2^z\sigma_4^z-\sigma_3^z\sigma_4
^z-h_4\sigma_4^x$$ and we wish to compute the magnetization $\langle \sigma_4^x\rangle$. More 
then that, we will compute the reduced state of site $4$. To do this, for the ease of 
computation, we first perform the partial trace over site $1$, followed by the partial trace of 
sites $2$ and $3$, giving the desired reduced state.

We can write the Hamiltonian as 
\begin{equation}
H=(H'\otimes \mathds{1}_4).(\mathds{1}_1\otimes H'')
\end{equation}
where $H'=-\sigma_1^z\sigma_2^z-\sigma_2^z\sigma_3^z-h_1\sigma_1^x$, defined on the space of sites $1$, $2$ and $3$, and $H''=-\sigma_2^z\sigma_4^z-\sigma_3^z\sigma_4
^z-h_4\sigma_4^x$, defined on the space of sites $2$, $3$ and $4$. Then we have that
\begin{equation}
e^{-\beta H}=(e^{-\beta H'}\otimes \mathds{1}_4).(\mathds{1}_1\otimes e^{-\beta H''}),
\end{equation}
so we get that
\begin{equation}
\text{Tr}_1(e^{-\beta H})=\{\text{Tr}_1(e^{-\beta H'})\otimes \mathds{1}_4\}.e^{-\beta H''}. \label{equaçao48}
\end{equation}
To calculate $\text{Tr}_1(e^{-\beta H'})$, let us compute $\text{Tr}_1(H'^n)$ and use the series expansion of $e^{-\beta H'}$ (\ref{eqtraco}) to find its partial trace. So, we have that
\begin{equation}
H'^2=(2+h_1^2)\mathds{1}+2\sigma_2^z\sigma_3^z=a\mathds{1}+2\sigma_2^z\sigma_3^z
\end{equation}
where $a=2+h_1^2$. Then, it is easy to find even powers of $H'$, that is
\begin{align}
H'^{2n}=\sum_{k=0}^n \binom{n}{k}a^{n-k}2^k(\sigma_2^z\sigma_3^z)^k
\end{align}
\begin{align}
=\left( \sum_{\overset{k=0}{even}}^n\binom{n}{k}a^{n-k}2^k \right)\mathds{1}+\left( \sum_{\overset{k=0}{odd}}^n\binom{n}{k}a^{n-k}2^k \right)\sigma_2^z\sigma_3^z. \nonumber
\end{align}
Summarizing, we can write
\begin{equation}
H'^{2n}=b_n \mathds{1} +c_n\sigma_2^z\sigma_3^z. \label{equation55}
\end{equation}
With this equation it is possible to calculate odd powers of $H'$ also, that is
\begin{equation*}
H'^{2n+1}=(b_n \mathds{1} +c_n\sigma_2^z\sigma_3^z).(\sigma_1^z\sigma_2^z+\sigma_2^z\sigma_3^z+h_1\sigma_1^x).
\end{equation*}
The above equation shows us that $\text{Tr}(H'^{2n+1})=0$, for all $n=0,1,2...$. Then we have that
\begin{align}
\text{Tr}_1(e^{-\beta H'})=\text{Tr}_1\left(\sum_{m=0}^\infty \frac{\beta^m}{m!}H'^m\right) \\
=\sum_{m=0}^\infty \frac{\beta^m}{m!}\text{Tr}_1(H'^m)=\sum_{\overset{m=0}{even}}^\infty 
\frac{\beta^m}{m!}\text{Tr}_1(H'^m).
\end{align}
Taking $m=2n$ and using Equation (\ref{equation55}), we have that
\begin{align}
\text{Tr}_1(e^{-\beta H'})=\sum_{n=0}^\infty \frac{\beta^{2n}}{(2n)!}\text{Tr}_1{H'^2n} \\
=\sum_{n=0}^\infty \frac{\beta^{2n}}{(2n)!}(2b_n\mathds{1}+2c_n\sigma_2^z\sigma_3^z),
\end{align}
and, after some arrangements, we get
\begin{align}
=\left(\sum_{n=0}^\infty \frac{\beta^{2n}2b_n}{(2n)!}\right)\mathds{1}+\left(\sum_{n=0}^\infty \frac{\beta^{2n}2c_n}{(2n)!}\right)\sigma_2^z\sigma_3^z.
\end{align}
We can write this as
\begin{equation}
\text{Tr}_1(e^{-\beta H'})=A\mathds{1}+B\sigma_2^z\sigma_3^z, \label{equacao61}
\end{equation}
where
\begin{equation}
A=\sum_{n=0}^\infty \sum_{\overset{k=0}{even}}^\infty \frac{2\beta^{2n}}{(2n)!}\frac{n!}{(n-k)!k!}a_1^{n-k}2^k
\end{equation}
and
\begin{equation}
B=\sum_{n=0}^\infty \sum_{\overset{k=0}{odd}}^\infty \frac{2\beta^{2n}}{(2n)!}\frac{n!}{(n-k)!k!}a_1^{n-k}2^k.
\end{equation}
Now, let us calculate $e^{-\beta H''}$. The powers of $H''$ will follow the same arguments of the 
powers of $H'$, so we have that
\begin{equation}
H''^{2n}=\alpha_{n}\mathds{1}+\epsilon_{n}\sigma_2^z\sigma_3^z
\end{equation}
and
\begin{align}
H''^{2n+1}=(\alpha_{n}\mathds{1}+\gamma_{n}\sigma_2^z\sigma_3^z)(\sigma_2^z\sigma_4^z+\sigma_3^z\sigma_3^z+h_4\sigma_4^x).
\end{align}
where $\alpha_n$ and $\epsilon_n$ can be found analogously to what was done for $b_n$ and $c_n$. 
Using the Taylor series of $e^{-\beta H''}$, we have that
\begin{align}
e^{-\beta H''}&=C\mathds{1}+D\sigma_2^z\sigma_3^z+E\sigma_2^z\sigma_4^z+F\sigma_3^z\sigma_4^z \nonumber \\
&+G\sigma_4^x+H\sigma_2^z\sigma_3^z\sigma_4^x \label{equacao66}
\end{align}
where
\begin{equation}
C=\sum_{n=0}^\infty \sum_{\overset{k=0}{even}}^\infty \frac{2\beta^{2n}}{(2n)!}\frac{n!}{(n-k)!k!}a_4^{n-k}2^k,
\end{equation}
\begin{equation}
D=\sum_{n=0}^\infty \sum_{\overset{k=0}{odd}}^\infty \frac{2\beta^{2n}}{(2n)!}\frac{n!}{(n-k)!k!}a_4^{n-k}2^k,
\end{equation}
\begin{equation}
G=h_4\sum_{n=0}^\infty \sum_{\overset{k=0}{even}}^\infty \frac{2\beta^{2n+1}}{(2n+1)!}\frac{n!}{(n-k)!k!}a_4^{n-k}2^k
\end{equation}
and
\begin{equation}
H=h_4\sum_{n=0}^\infty \sum_{\overset{k=0}{odd}}^\infty \frac{2\beta^{2n+1}}{(2n+1)!}\frac{n!}{(n-k)!k!}a_4^{n-k}2^k,
\end{equation}
where we have set $a_4=2+h_4^2$. We do not show the expressions of $E$ and $F$ here, since 
they are not used.

Putting Equations (\ref{equacao61}) and (\ref{equacao66}) in Equation \ref{equaçao48}, performing 
the partial trace on spaces of sites $2$ and $3$, and normalizing the trace of resulting operator, 
we have that the reduced state of site $4$ is given by
\begin{equation}
\rho_4=\frac{1}{2}\mathds{1}+\frac{1}{2}\frac{AG+BH}{AC+BD}\sigma_4^x.
\end{equation}
Finally, we get that
\begin{equation}
\langle \sigma_4^x \rangle=\frac{AG+BH}{AC+BD}. \label{magnetizaçao}
\end{equation}

\begin{figure}
\includegraphics[scale=0.19]{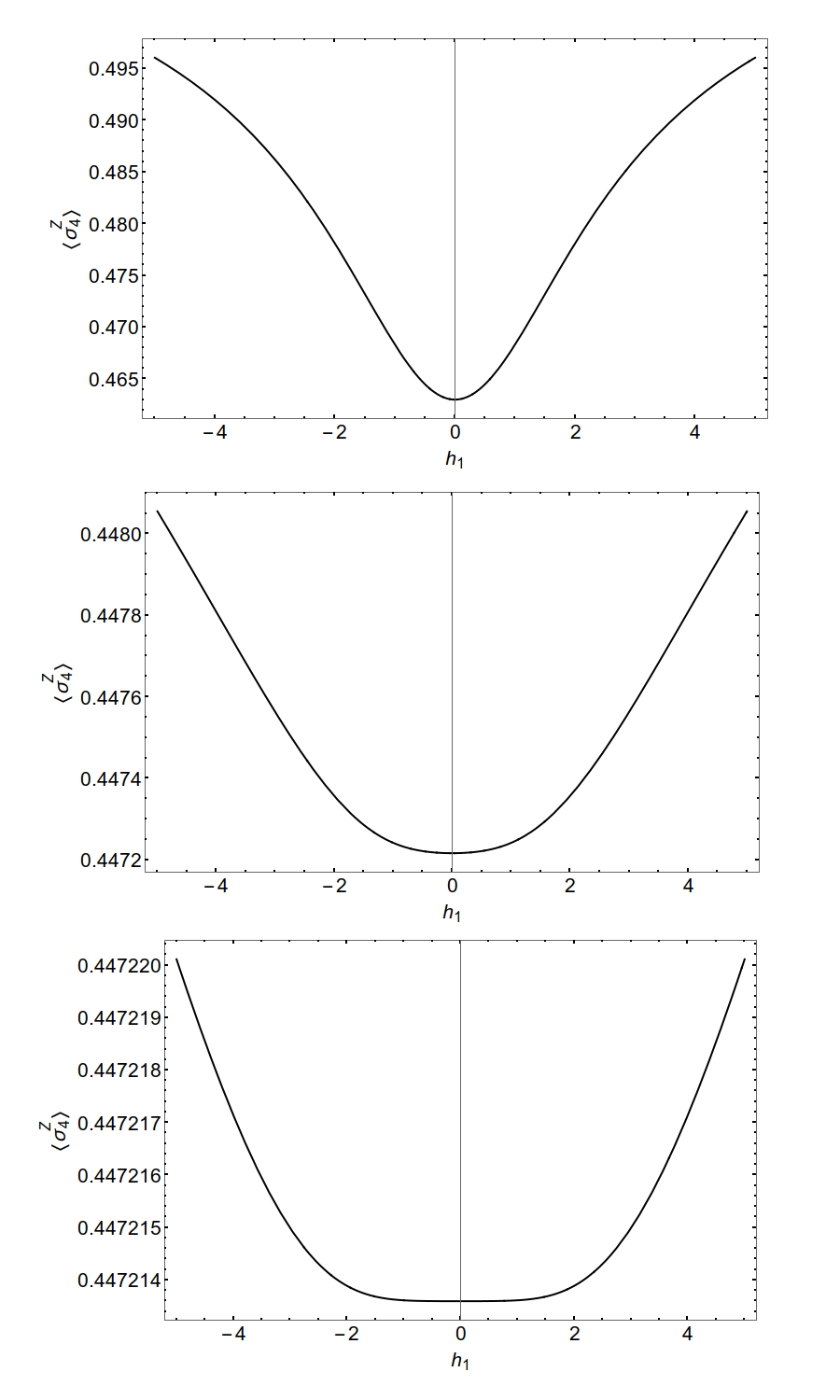}
\caption{Graphics of the value of the magnetization $\langle \sigma_4^x \rangle$ of site $4$ in function of the external magnetic field $h_1$ applied in site $1$. Each graphic was done for one different value of $\beta$, which are $\beta=1$, $4$ and $7$. Note that the scale of each graphic is different.} \label{graficomag}
\end{figure}

We have calculated the expressions of the series that define coefficients $A,B,C,D,G$ and $H$ which 
are analytical expressions. The graphics of the magnetization \ref{magnetizaçao} in function of 
$h_1$ are given in Figure \ref{graficomag} for some values of $\beta$, which are $\beta=1$, $4$ and 
$7$ (note that each graphic has a different scale). Furthermore, we can show that the magnetization 
(\ref{magnetizaçao}) is independent of $h_1$ when the value of $\beta$ goes to infinity. More 
specifically, it is equal to $\frac{1}{\sqrt{4+h_4^2}}$ when $\beta\rightarrow\infty$.

In conclusion, it is shown that the shielding property do not work, in general, when the interface 
has more than one site when the temperature is positive. For null temperature, however, the 
shielding property still works in this example.


\begin{thebibliography}{99}


\bibitem{LSM} E. H. Lieb, T. D. Schultz, and D. C. Mattis,Two soluble models of an antiferromagnetic chain, Ann. Phys. (N. Y.)16, 407 (1961).
\bibitem{IPT} B. K. Chakrabarti, A. Dutta, and P. Sen,Quantum Ising Phases and Transitions in Transverse Ising models(Springer-Verlag, Berlin, 1996).
\bibitem{ATec}T. Giamarchi, Quantum Physics in One Dimension (Oxford University Press, Oxford, 2004).
\bibitem{Ntec} F. Verstraete, D. Porras, J. I. Cirac, DMRG and periodic boundary conditions: a quantum information perspective, Phys. Rev. Lett. 93, 227205 (2004); R. Orus, A Practical Introduction to Tensor Networks: Matrix Product States and Projected Entangled Pair States, Annals of Physics 349 (2014) 117-158.
\bibitem{EPT} Fernando G. S. L. Brand\~ao, Entanglement as order parameter, New J. Phys. 7, 254 (2005).
\bibitem{DOS} P. Haikka, J. Goold, S. McEndoo, F. Plastina, S. Maniscalco, Non-Markovianity, Loschmidt echo and criticality: a unified picture, Phys. Rev. A 85, 060101(R) (2012).
\bibitem{QTD} F. Cosco, M. Borrelli, P. Silvi, S. Maniscalco, G. De Chiara, Non-equilibrium quantum thermodynamics in Coulomb crystals, Phys. Rev. A 95, 063615 (2017); 

L. Fusco, S. Pigeon, T. J. G. Apollaro, A. Xuereb, L. Mazzola, M. Campisi, A. Ferraro, M. Paternostro, G. De Chiara, Assessing the non-equilibrium thermodynamics in a quenched quantum many-body system via single projective measurements, Phys. Rev. X 4, 031029 (2014).
\bibitem{cold} J. Simon, W.m S. Bakr, R. Ma, M. E. Tai, P. M. Preiss, M. Greiner, Quantum Simulation of Antiferromagnetic Spin Chains in an Optical Lattice, Nature 472, 307(2011).
\bibitem{solid} R. Coldea, D.A. Tennant, E.M. Wheeler, E. Wawrzynska, D. Prabhakaran, M. Telling, K. Habicht, P. Smeibidl, K. Kiefer, Quantum criticality in an Ising chain: experimental evidence for emergent E8
symmetry, Science 327, 177 (2010).
\bibitem{duality} P. Fendley, Modern Statistical Mechanics, http://galileo.phys.virginia.edu/~pf7a/book.html, in preparation.
\bibitem{roeck} W. De Roeck, M. Schutz, Local Perturbations Perturb -Exponentially- Locally, J. Math. Phys. 56, 061901 (2015).
\bibitem{sven} S. Bachmann, S. Michalakis, B. Nachtergaele, R. Sims, Automorphic Equivalence within Gapped Phases of Quantum Lattice Systems, Commun. Math. Phys. 309, 835-871 (2012).
\bibitem{LR}E. Lieb, D. Robinson, The Finite Group Velocity of Quantum Spin Systems, Commun. Math. 
Phys. 28, 251-257 (1972).  
\bibitem{lightcone} R. C. Drumond, N. S. M\'oller, Bounding entanglement spreading after a local quench,  	Physical Review A 95, 062301 (2017).
\bibitem{Bravyi} S. Bravyi, M. B. Hastings, F. Verstraete, Lieb-Robinson bounds and the generation of correlations and 
topological quantum order, Phys. Rev. Lett. 97, 050401 (2006). 


\end{thebibliography}
\end{document}